\documentclass[11pt]{article}
\usepackage[english]{babel}
\usepackage{amsmath} \usepackage{amssymb} \usepackage{bm}
\usepackage{graphicx}
\RequirePackage[numbers,sort&compress]{natbib}
\usepackage{color}

\def\be{\begin{equation}}
\def\ee{\end{equation}}

\begin{document}

\begin{center}
{\Large \bf On the spacetime connecting two aeons}
\vskip 0.1cm
{\Large \bf in conformal cyclic cosmology}
\vskip 0.5cm
{\bf A. Araujo, H. Jennen, J. G. Pereira, A. C. Sampson and L. L. Savi}\footnote{Current address: Universidade Tecnol\'ogica Federal do Paran\'a, C\^ampus Pato Branco, Via do Conhecimento Km 1, Pato Branco, PR, 85503-390, Brazil.}
\vskip 0.1cm
{\it Instituto de F\'{\i}sica Te\'orica, Universidade Estadual Paulista \\
Rua Dr. Bento Teobaldo Ferraz 271 \\
01140-070 S\~ao Paulo, Brazil}
\end{center}

\vskip 0.3cm
\centerline{\bf Abstract}
\begin{quote}
{\footnotesize As quotient spaces, Minkowski and de Sitter are fundamental, non-gravitational spacetimes for the construction of physical theories. When general relativity is constructed on a de Sitter spacetime, the usual Riemannian structure is replaced by a more general structure called de Sitter-Cartan geometry. In the contraction limit of an infinite cosmological term, the de Sitter-Cartan spacetime reduces to a singular, flat, conformal invariant four-dimensional cone spacetime, in which our ordinary notions of time interval and space distance are absent. It is shown that such spacetime satisfies all properties, including the Weyl curvature hypothesis, necessary to play the role of the bridging spacetime connecting two aeons in Penrose's conformal cyclic cosmology.

}
\end{quote}

\section{Introduction}

A possible interpretation of the Planck length is that it represents the minimum attainable length in nature. If this comes to be true, it would emerge as an invariant length parameter, whose existence would modify the physics of the Planck scale. For example, the spacetime local kinematics at that scale would no longer be described by ordinary special relativity because, as is well-known, it does not allow the existence of such invariant scale. One should then look for a modified special relativity. An interesting attempt in this direction is the so-called ``doubly special relativity'' \cite{dsr1,dsr2}, obtained by introducing into the dispersion relation of special relativity scale-suppressed terms of higher order in the momentum, in such a way to allow the existence of an invariant length at the Planck scale. The importance of these terms is controlled by a parameter $\kappa$, which changes the kinematic group of special relativity from Poincar\'e to a $\kappa$-deformed Poincar\'e group, in which Lorentz symmetry is explicitly violated. Far away from the Planck scale these terms are suppressed, Lorentz symmetry is recovered and one obtains back ordinary special relativity.

A different solution to the same problem shows up from noting that {\em Lorentz transformations do not change the curvature of the homogeneous spacetime in which they are performed} \cite{dSsr0,dSsr1,dSsr2}. Considering that the scalar curvature $R$ of any homogeneous spacetime is of the form
\[
R \sim \pm \, l^{-2},
\]
with $l$ its pseudo-radius, {Lorentz transformations are then found to leave the length parameter $l$ invariant}. Although somewhat hidden in Minkowski space, because what is left invariant in this case is an infinite length---corresponding to a vanishing scalar curvature---in de Sitter and anti-de Sitter spacetimes, whose pseudo-radii are finite, this property becomes manifest. One then sees that, contrary to the usual belief, {\em Lorentz transformations do leave invariant a very particular length: that defining the scalar curvature of the homogeneous spacetime}.

If the Planck length $l_P$ is to be invariant under Lorentz transformations, therefore, it is natural to assume that it represents the pseudo-radius of spacetime at the Planck scale, which will then be either a de Sitter or an anti-de Sitter space, with the scalar curvature given by 
\be
R \sim \pm \, l_P^{-2} \simeq \pm \, 10^{66}\, {\rm cm}^{-2},
\label{PlanLam}
\ee
where the $+$ ($-$) sign refers to the de Sitter (anti de Sitter) case. From now on, taking into account recent astronomical observations indicating that the universe expansion is presently accelerating, we will restrict ourselves to the de Sitter case.

Now, as quotient spaces, Minkowski and de Sitter represent different non-gravitational backgrounds for the construction of physical theories. General relativity, for instance, can be constructed on any one of them. Of course, in either case gravitation will have the same dynamics, only their local kinematics will be different. If the underlying spacetime is Minkowski, the local kinematics will be ruled by the Poincar\'e group of ordinary special relativity. If the underlying spacetime is de Sitter, the local kinematics will be ruled by the de Sitter group, which amounts then to replace ordinary special relativity by a de Sitter-ruled special relativity. It is important to remark that, even though there is an invariant length-parameter related to the cosmological term, the Lorentz group remains part of the spacetime kinematics, which means that this symmetry is not broken at any energy scale. Taking into account the deep connection between Lorentz symmetry and causality \cite{zeeman}, this theory predicts that causality is always preserved, even at the Planck scale.

When general relativity is constructed on de Sitter, spacetime will no longer present a Riemannian structure, but will be described by a particular case of a more general structure called Cartan geometry \cite{CartanGeo}. As a matter of fact, it will be described by a Cartan geometry that reduces locally to de Sitter---and for this reason is called de Sitter-Cartan geometry \cite{wise}. Accordingly, in a locally inertial frame, where inertial effects exactly compensate for gravitation, the spacetime metric reduces to the de Sitter metric.
One should note that such construction does not change the dynamics of the gravitational field, which remains described by Einstein equation. The only change will be in the strong equivalence principle, which passes to state that {\em in a locally inertial frame, where gravitation goes unnoticed, the laws of physics reduce to those of de Sitter-ruled special relativity.}

Considering a general de Sitter-Cartan spacetime, and making use of the In\"on\"u-Wigner process of group and algebra contractions \cite{inonu1}, the purpose of this paper is to study
both the contraction limit for the pseudo-radius going to infinity $l \to \infty$, which corresponds to a vanishing cosmological term $\Lambda \to 0$, and the contraction limit for a vanishing pseudo-radius $l \to 0$, which corresponds to an infinite cosmological term $\Lambda \to \infty$. In the first case, the underlying de Sitter spacetime contracts to Minkowski, and the de Sitter-Cartan geometry reduces to the usual Riemannian geometry of general relativity. In the second case, the gravitational degrees of freedom are switched off, and the whole de Sitter-Cartan spacetime contracts to a kinematic four-dimensional homogeneous, conic spacetime, which is singular at the origin. Such spacetime has already been shown to bear algebraic, geometric and thermodynamic properties that fit quite reasonably to what one would expect for an initial condition of a big bang universe \cite{cone}. In this paper we are going to show that it meets also the properties required for playing the role of the bridging spacetime connecting two aeons in Penrose's conformal cyclic cosmology \cite{PenBook}.

\section{The de Sitter spacetime and group: possible limits}

The maximally symmetric de Sitter spacetime, denoted $dS$, can be seen as a hypersurface in the host pseudo-Euclidean space with metric $\eta_{AB}$ = $(+1,-1,-1,-1,-1)$ ($A, B, ... = 0, \dots, 4)$, whose points in Cartesian coordinates $\chi^A$ satisfy the relation \cite{ellis}
\be
\eta_{AB} \, \chi^A \chi^B = - \, l^2,
\label{AmbiCoord}
\ee
or equivalently, in four-dimensional coordinates,
\be
\eta_{\mu \nu} \, \chi^{\mu} \chi^{\nu} - 
\big(\chi^4 \big)^2 = - \, l^2.
\label{dspace2}
\ee
It has the de Sitter group $SO(4,1)$ as group of motions, and is homogeneous under the Lorentz group ${\mathcal L} = SO(3,1)$, that is \cite{kono}
\be
dS = SO(4,1) / {\mathcal L}.
\label{quotient}
\ee
In the ambient Cartesian coordinates $\chi^A$, the generators of the infinitesimal de Sitter transformations are written in the form
\be
L_{A B} = \eta_{AC} \, \chi^C \, \frac{\partial}{\partial \chi^B} 
-
\eta_{BC} \, \chi^C \, \frac{\partial}{\partial \chi^A}.
\label{dsgene}
\ee
They satisfy the commutation relations
\be
\left[ L_{AB}, L_{CD} \right] = \eta_{BC} L_{AD} + \eta_{AD} 
L_{BC} - \eta_{BD} L_{AC}
- \eta_{AC} L_{BD}.
\label{dsal}
\ee

On account of the quotient character (\ref{quotient}) of the de Sitter spacetime, geometry and algebra turns out to be deeply connected. As a consequence, any deformation in the algebra and group will produce concomitant deformations in the geometry of the corresponding homogeneous spacetime. The use of the In\"on\"u-Wigner contraction, therefore, constitutes a reliable method for studying geometrical limits of homogeneous spacetimes. In what follows we consider first, in the guise of completeness, the well-known contraction limit $l \to \infty$. Subsequently we consider the contraction limit $l \to 0$, whose study constitutes the main part of the paper. Since these two limits require a different parameterisation, they must be performed separately.

\subsection{Large pseudo-radius contraction}

\subsubsection{Parameterisation appropriate for large values of $l$}

The four-dimensional stereographic coordinates $\{x^\mu\}$ are obtained through a ste\-re\-o\-graphic projection from the de Sitter hypersurface into a target Minkowski spacetime. In the parameterisation appropriate to deal with large values of $l$, they are defined by \cite{gursey}
\be
\chi^{\mu} = \Omega \, x^\mu
\label{xix}
\ee
and
\be
\chi^4 = - \, \Omega \left(1 + 
{\sigma^2}/{4 l^2} \right)
\label{xi4}
\ee
where $\Omega \equiv \Omega(x)$ is the function
\be
\Omega = (1 - {\sigma^2}/{4 l^2})^{-1}
\label{n}
\ee
with $\sigma^2$ the Lorentz invariant quadratic form $\sigma^2 = \eta_{\mu \nu} \, x^\mu x^\nu$. In these coordinates, the infinitesimal de Sitter quadratic interval
\be
ds^2 = g_{\alpha \beta} \, dx^\alpha dx^\beta
\ee
is written with the conformally flat metric
\be
g_{\alpha \beta} = \Omega^{2} \, \eta_{\alpha \beta}.
\label{44Sbis}
\ee
The corresponding Christoffel connection is \cite{livro}
\be
\Gamma^{\lambda}{}_{\mu \nu} = \frac{\Omega}{2 l^2} \big( \delta^{\lambda}_{\mu} \,
\eta_{\nu \alpha} \, x^\alpha  + \delta^{\lambda}_{\nu}\,
\eta_{\mu \alpha} \, x^\alpha - \eta_{\mu \nu} \, x^\lambda \big)
\label{46}
\ee
with the Riemann tensor given by
\be
R^{\mu}{}_{\nu \rho \sigma} = \frac{\Omega^2}{l^2}
\left(\delta^{\mu}_{\rho} \, \eta_{\nu \sigma} -
\delta^{\mu}_{\sigma} \, \eta_{\nu \rho} \right).
\label{47}
\ee
The Ricci and the scalar curvature are, consequently,
\be
R_{\nu \sigma} = \frac{3 \Omega^2}{l^2} \, \eta_{\nu \sigma}
\qquad \mbox{and} \qquad R = \frac{12}{l^2}.
\label{RiciScalar}
\ee

In terms of stereographic coordinates $\{x^\mu\}$, the de Sitter generators~(\ref{dsgene}) 
are written in the form
\be
L_{\mu \nu} =
\eta_{\mu \rho} \, x^\rho \, P_\nu - \eta_{\nu \rho} \, x^\rho \, P_\mu
\label{dslore}
\ee
and
\be
L_{4 \mu} = l \, P_\mu - \frac{1}{4 l} 	\, K_\mu
\label{dstra}
\ee
where
\be
P_\mu = \partial_\mu \quad {\rm and} \quad
K_\mu = \left(2 \eta_{\mu \nu} x^\nu x^\rho - \sigma^2 
	\delta_{\mu}^{\rho}
\right) \partial_\rho
\label{cp2}
\ee
are, respectively, the generators of translations and {proper} conformal transformations \cite{coleman}. Generators $L_{\mu \nu}$ refer to the Lorentz subgroup, whereas the elements $L_{4 \mu}$ define the transitivity on the homogeneous space.\footnote{See the Appendix for a mathematical definition of transitivity of homogeneous spaces.} From Eq.~(\ref{dstra}) it follows that the de Sitter spacetime is transitive under a combination of translations and proper conformal transformations --- usually called de Sitter ``translations''. The relative importance between these two transformations is determined by the value of the pseudo-radius $l$.

In order to study the limit of large values of $l$, it is necessary to parameterise the generators (\ref{dstra}) according to \cite{gursey}
\be
\Pi_\mu \equiv \frac{L_{4 \mu}}{l} = P_\mu - 
	\frac{1}{4 l^2} \, K_\mu.
\label{TransiS}
\ee
In terms of these generators, the de Sitter algebra (\ref{dsal}) assumes the form
\begin{align}
\left[L_{\mu \nu}, L_{\rho \sigma}\right] & = \eta_{\nu \rho} \,
L_{\mu \sigma} + \eta_{\mu \sigma} \, L_{\nu \rho} - \eta_{\nu 
	\sigma} \,
L_{\mu \rho} - \eta_{\mu \rho} \, L_{\nu \sigma}, \label{dS1a}  \\
\left[ \Pi_{\mu}, L_{\rho \sigma}\right]& = \eta_{\mu \rho}  
\Pi_{\sigma} -
\eta_{\mu \sigma}  \Pi_{\rho}, \label{dS2a} \\
\left[ \Pi_{\mu}, \Pi_{\rho}\right]& = l^{-2} 
L_{\mu \rho}. \label{dS3a}
\end{align}
The last commutator shows that the de Sitter ``translation'' generators are not really translations, but rotations; hence the quotation marks.

\subsubsection{The contraction limit $l \to \infty$}

In the limit $l \to \infty$, we see from Eq.~(\ref{TransiS}) that the de Sitter generators $\Pi_\mu$ reduce to generators of ordinary translations
\be
\Pi_\mu \to P_\mu.
\label{PtoPi}
\ee
Concomitantly, the de Sitter algebra (\ref{dS1a}-\ref{dS3a}) contracts to
\begin{align}
\left[L_{\mu \nu}, L_{\rho \sigma}\right] &= \eta_{\nu \rho} \,
L_{\mu \sigma} + \eta_{\mu \sigma} \, L_{\nu \rho} - \eta_{\nu 
\sigma} \, L_{\mu \rho} - \eta_{\mu \rho} \, L_{\nu \sigma}   \\
\left[ P_{\mu}, L_{\rho \sigma}\right]&=\eta_{\mu \rho} 
P_{\sigma} - \eta_{\mu \sigma} P_{\rho}  \\
\left[ P_{\mu}, P_{\rho}\right]&= 0  
\end{align}
which is the Lie algebra of the Poincar\'e group ${\mathcal P} = {\mathcal L} \, 
\oslash \, {\mathcal T}$, the semi-direct product of the Lorentz 
(${\mathcal L}$) and the translation (${\mathcal T}$) groups.
As a result of this algebra and group deformations, the de Sitter spacetime $dS$ contracts to the flat Minkowski space $M$:
\be
dS \to M = {\mathcal P}/{\mathcal L}.
\label{dStoM}
\ee
In fact, as a simple inspection shows, the de Sitter metric (\ref{44Sbis}) reduces to the Minkowski metric
\be
g_{\mu \nu} \to \eta_{\mu \nu},
\ee
and the Riemann, Ricci and scalar curvatures vanish identically:
\be
{R}^{\mu}{}_{\nu \rho \sigma} \to 0, \quad 
{R}_{\nu \sigma} \to 0, \quad 
{R} \to 0.
\ee
From (\ref{PtoPi}) we see that Minkowski spacetime $M$ is {transitive} under ordinary translations: the point-set of $M$ is determined by ordinary translations.

\subsection{Small pseudo-radius contraction}

We proceed now to study the contraction limit of a small de Sitter pseudo-radius, which corresponds to an infinite cosmological term. This is the limit we are most interested in this paper.

\subsubsection{Parameterisation appropriate for small values of $l$}

To deal with small values of $l$, it is convenient to define the `inverse' host space coordinates
\be
\bar{\chi}^A = \chi^A / 4 l^2,
\label{BarCoord}
\ee
in terms of which relation (\ref{dspace2}) assumes the form
\be
\eta_{\mu \nu} \, \bar \chi^{\mu} \bar \chi^{\nu} - 
\left(\bar \chi^4 \right)^2 = - \, \frac{1}{16 l^2}.
\label{dspace2bisbis}
\ee
The stereographic projection is now defined by
\be
\bar{\chi}^\mu = \bar{\Omega} \, x^\mu
\label{chibarster}
\ee
and
\be
\bar \chi^4 = - \, l \, \bar \Omega \left(1 + {\sigma^2}/{4 l^2} \right)
\label{Barxi4}
\ee
where
\be
\bar{\Omega} \equiv \frac{\Omega}{4l^2} = \frac{1}{4l^2 - \sigma^2}.
\label{n3bar}
\ee
In these coordinates, the infinitesimal de Sitter quadratic interval
\be
d\bar s^2 = \bar{g}_{\alpha \beta} \, dx^\alpha dx^\beta
\ee
is written with the metric
\be
\bar{g}_{\alpha \beta} = \bar{\Omega}^{2} \, \eta_{\alpha \beta}.
\label{44bis2}
\ee

Considering that $\bar{g}_{\alpha \beta}$ and ${g}_{\alpha \beta}$ differ by a constant, the corresponding Christoffel connections will coincide:
\be
\bar{\Gamma}^{\lambda}{}_{\mu \nu} \equiv {\Gamma}^{\lambda}{}_{\mu \nu} =
2 \bar{\Omega} \, \big( \delta^{\lambda}_{\mu} \,
\eta_{\nu \alpha} \, x^\alpha  + \delta^{\lambda}_{\nu}\,
\eta_{\mu \alpha} \, x^\alpha - \eta_{\mu \nu} \, x^\lambda \big).
\label{46bis}
\ee
Of course, the same happens to the Riemann tensor
\be
\bar{R}^{\mu}{}_{\nu \rho \sigma} \equiv {R}^{\mu}{}_{\nu \rho \sigma} =
16 \, {l^2} \, \bar{\Omega}^2
\left(\delta^{\mu}_{\rho} \, \eta_{\nu \sigma} -
\delta^{\mu}_{\sigma} \, \eta_{\nu \rho} \right), 
\label{47bis}
\ee
as well as to the Ricci tensor:
\be
\bar{R}_{\nu \sigma} \equiv {R}_{\nu \sigma} =
16 \, {l^2} \, \bar{\Omega}^2 \, \eta_{\nu \sigma}.
\label{RiciBar}
\ee
The scalar curvature, however, due to a further contraction with the metric tensor, assumes a different form
\be
\bar{R} \equiv 16 \, l^4 R = 192 \, {l^2}.
\label{ScalarBar}
\ee
Considering that the cosmological term is proportional to
\be
\Lambda \sim l^{- 2},
\ee
it relates to the scalar curvature according to
\be
\Lambda \sim \frac{\bar R}{l^4}.
\ee

The de Sitter generators (\ref{dslore}) are the same for both parameterisations. However, in order to study the limit of small values of $l$, it is necessary to rewrite generators (\ref{dstra}) in the form \cite{cone}
\be
\bar{\Pi}_\mu \equiv 4l \, L_{4 \mu} = 4 l^2 P_\mu - K_\mu,
\label{TransGenL}
\ee
in terms of which the de Sitter algebra (\ref{dsal}) becomes
\begin{align}
\left[L_{\mu \nu}, L_{\rho \sigma}\right] &= \eta_{\nu \rho} \,
L_{\mu \sigma} + \eta_{\mu \sigma} \, L_{\nu \rho} - \eta_{\nu 
	\sigma} \,
L_{\mu \rho} - \eta_{\mu \rho} \, L_{\nu \sigma}  \\
\left[ \bar{\Pi}_{\mu}, L_{\rho \sigma}\right]&=\eta_{\mu \rho}  
\bar{\Pi}_{\sigma} -
\eta_{\mu \sigma}  \bar{\Pi}_{\rho}   \\
\left[ \bar{\Pi}_{\mu}, \bar{\Pi}_{\rho}\right]& = 16 \, l^{2} L_{\mu \rho}. 
\end{align}

\subsubsection{The contraction limit $l \to 0$}
\label{l0contra}

In the contraction limit $l \rightarrow 0$, the generators $\bar{\Pi}_\mu$ reduce to (minus) the proper conformal generators: 
\be
\bar{\Pi}_\mu \to  - K_\mu.
\ee
Accordingly, the de Sitter group $SO(4,1)$ contracts to the {\em conformal} Poincar\'e group \cite{ap1}
\[
SO(4,1) \to \bar{\mathcal P} = {\mathcal L} \oslash  \bar{\mathcal T},
\]
the semi-direct product between the Lorentz ${\mathcal L}$ and the proper conformal group $\bar{\mathcal T}$, whose Lie algebra is
\begin{align}
\left[{L}_{\mu \nu}, {L}_{\lambda \rho}\right] &= {\eta}_{\nu \lambda} \,
{L}_{\mu \rho} + {\eta}_{\mu \rho} \, {L}_{\nu \lambda} - {\eta}_{\nu \rho} \,
{L}_{\mu \lambda} - {\eta}_{\mu \lambda} \, {L}_{\nu \rho} \label{llx} \\
\left[ K_{\mu}, L_{\lambda \rho}\right]&= {\eta}_{\mu \lambda}  K_{\rho} -
{\eta}_{\mu \rho}  K_{\lambda} \label{klx} \\
\left[ K_{\mu},  K_{\lambda}\right]&= 0 \, .
\end{align}
The name of this group stems from the fact that it has the same Lie algebra of the ordinary Poincar\'e group, but with the translation generators $P_\mu$ replaced by the proper conformal generators $K_\mu$. Concomitant with the group contraction, on account of their quotient character, the de Sitter spacetime $dS$ reduces to the homogeneous space $\bar M$
\be
dS \to \bar{M} = \bar{\mathcal P}/{\mathcal L}.
\ee
The conformal Poincar\'e group $\bar{\mathcal P}$, like the Poincar\'e group ${\mathcal P}$, has the Lorentz group ${\mathcal L}$ as the subgroup accounting for the isotropy of the space. The homogeneity, however, is completely different: instead of ordinary translations, all points of $\bar{M}$ are equivalent under special conformal transformations. In other words, the point-set of $\bar{M}$ is that determined by special conformal transformations.

\begin{figure}[ht]
\begin{center}
\scalebox{0.45}{\includegraphics{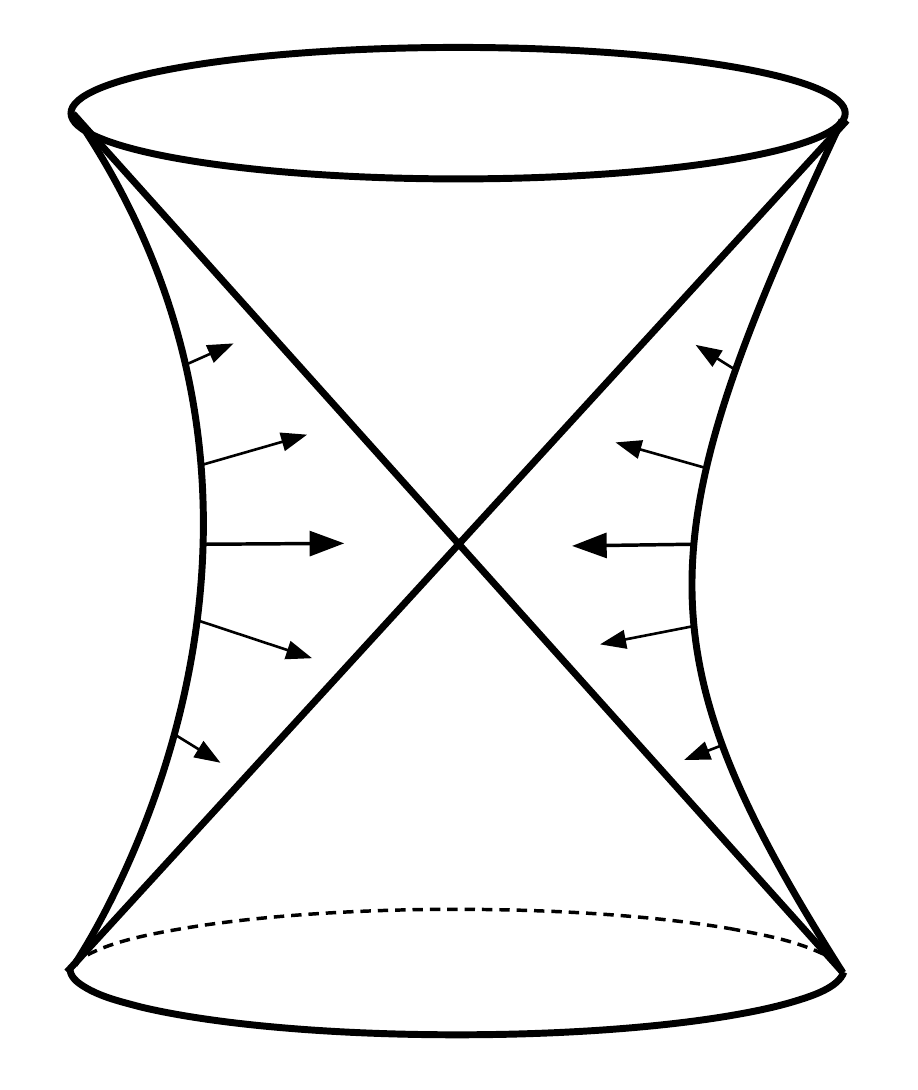}}
\caption{\it Pictorial view of the de Sitter spacetime in the contraction limit $l \to 0$, showing its deformation into a four-dimensional cone spacetime.} 
\label{Fig1}
\end{center}
\end{figure}

In the limit $l \to 0$, the de Sitter metric (\ref{44bis2}) assumes the conformal invariant form \cite{TodGRG}
\be
\bar{g}_{\mu \nu} \to \bar{\eta}_{\mu \nu} = \sigma^{-4} \, \eta_{\mu \nu},
\label{ConeMetric}
\ee
which is the metric on $\bar{M}$. The Christoffel connection, on the other hand, reduces to
\be
{\bar \Gamma}^{\lambda}{}_{\mu \nu} \to {^0}{\bar \Gamma}^{\lambda}{}_{\mu \nu} = - \, 2 \sigma^{-2} \,\big( \delta^{\lambda}_{\mu} \,
\eta_{\nu \alpha} \, x^\alpha  + \delta^{\lambda}_{\nu} \,
\eta_{\mu \alpha} \, x^\alpha - \eta_{\mu \nu} \, x^\lambda \big).
\label{46bis0}
\ee
The corresponding Riemann, Ricci, and scalar curvatures vanish identically:
\be
{\bar{R}}^{\mu}{}_{\nu \rho \sigma} \to {^0}{\bar{R}}^{\mu}{}_{\nu \rho \sigma} = 0, \quad 
\bar{R}_{\nu \sigma} \to {^0}{\bar{R}}_{\nu \sigma} = 0, \quad 
\bar{R} \to {^0}{\bar{R}} = 0.
\ee
The cosmological term, however, goes to infinity:
\be
\Lambda \to \infty.
\ee
From these properties one can infer that $\bar{M}$ is a singular, four-dimensional cone spacetime (see Figure~\ref{Fig1}), transitive under proper conformal transformations \cite{cone}.
It is important to remark that the contraction limit $l \to 0$ is not continuous --- actually a general property of any group contraction. This can be seen by observing that, whereas the de Sitter group is semi-simple, the conformal Poincar\'e group is not, which means that it is not possible to continuously deform the former into the latter. It is this singular character that produces a decoupling between the scalar curvature and the cosmological term, making the former going to zero and the latter to infinity.

\section{Some attributes of the cone spacetime $\bar M$}
\label{attributes}

As is well-known, in $(3 + 1)$ dimensions there are three homogeneous spaces: Minkowski, de Sitter and anti-de Sitter. The cone spacetime $\bar{M}$ is an additional four-dimensional maximally symmetric spacetime, with the conformal Poincar\'e group $\bar{\mathcal P}$ as kinematic group, which is however singular. In this section we explore some of its properties.

\subsection{Geometric relation between $M$ and $\bar M$}

Under the spacetime inversion
\be
x^\mu \to -\,  \frac{x^\mu}{\sigma^2}
\label{inversion}
\ee
the translation generators are led to the proper conformal transformations, and vice versa \cite{coleman}:
\be
P_\mu \to K_\mu \qquad \mbox{and} \qquad K_\mu \to P_\mu.
\ee
The Lorentz generators, on the other hand, remain unchanged:
\be
L_{\mu \nu} \to L_{\mu \nu}.
\ee
This means that, under such inversion, the Poincar\'e group ${\mathcal P}$ is led to the conformal Poincar\'e group $\bar{\mathcal P}$, and vice versa. Concomitantly, Minkowski $M$ is transformed into the four-dimensional cone-spacetime $\bar{M}$, and vice versa. The corresponding spacetime metrics are also transformed into each other:
\be
\eta_{\mu \nu} ~\to~ \bar{\eta}_{\mu \nu} = \sigma^{-4} \, \eta_{\mu \nu} \qquad \mbox{and} \qquad
\bar{\eta}_{\mu \nu} = \sigma^{-4} \, \eta_{\mu \nu} ~\to~ \eta_{\mu \nu}.
\ee
Minkowski and the cone spacetimes can thus be considered a kind of {\it dual} to each other in the sense that their geometries are determined, respectively, by a vanishing and an infinite cosmological term. In the same way the Minkowski metric $\eta_{\mu \nu}$ is invariant under spacetime translations, the conic spacetime metric $\bar{\eta}_{\mu \nu}$ is invariant under proper conformal transformations.

\subsection{The nature of the spacetime singularity}

As we have already seen, the cone spacetime $\bar{M}$ has vanishing Riemann, Ricci and scalar curvature tensors. The cosmological term, on the other hand, is infinite, pointing to a singular spacetime. In fact, its metric tensor $\bar{\eta}_{\mu \nu}$, given by Eq.~(\ref{ConeMetric}), is singular at the vertex of the cone spacetime, located at $x^\mu = 0$, in which $\sigma^2 \equiv \eta_{\mu \nu} x^\mu x^\nu = 0$.\footnote{It is important to note that $\eta_{\mu \nu} x^\mu x^\nu \neq 0$ in all other points of the cone. What vanishes in all points of the cone is the quadratic form written in terms of the five-dimensional ambient space coordinates $\eta_{AB} \chi^A \chi^B = 0$, as follows from (\ref{AmbiCoord}) in the contraction limit $l \to 0$.} However, if we perform a conformal re-scaling of the metric
\be
\bar{\eta}_{\mu \nu} ~\to~ {\bar{\eta}}'_{\mu \nu} = \omega^2(x) \, \bar{\eta}_{\mu \nu},
\ee
with the conformal factor given by
\be
\omega^2(x) = \sigma^4 \, \alpha^2(x),
\ee
the resulting metric tensor
\be
{\bar{\eta}}'_{\mu \nu} = \alpha^2(x) \, \bar{\eta}_{\mu \nu}
\ee
is no longer singular. This kind of singularity, in which the metric is singular but the conformal equivalence class of the metric is not, is called a {\em conformal gauge singularity} \cite{tod0}. 

\subsection{Weyl curvature hypothesis}

The conformal gauge singularity is an instrumental part of the Weyl curvature hypothesis. In its original version, the hypothesis states that {\em the Weyl curvature must vanish at any initial singularity} \cite{PenCente}. Of course, as a flat spacetime, the cone spacetime naturally satisfies the Weyl curvature hypothesis. Afterwards, a new version of the hypothesis was put forward by P. Tod, which says that {\em every conformal singularity can be transformed into a smooth spacelike hypersurface by a conformal rescaling in such a way that geodesics can be extended beyond it.} \cite{tod}. Since the cone singularity is a conformal gauge singularity, it satisfies also Tod's version of the Weyl curvature hypothesis.

\section{Transitivity and the notion of distance and time}
\label{SpaceTime}

As already discussed, when general relativity is constructed on de Sitter, spacetime will no longer present a Riemannian structure, but will be described by a more general structure called de Sitter-Cartan geometry. The name stems from the fact that, in a locally inertial frame, where inertial effects exactly compensate for gravitation, the spacetime metric reduces to the (non-gravitational) de Sitter metric. Since any two points of de Sitter spacetime are connected to each other by a combination of translation and proper conformal transformations, the same happens locally in a de Sitter-Cartan spacetime. In the parameterisation appropriate for large values of $l$, any two points separated by an infinitesimal distance will be connected to each other by a transformation generated by
\be
\Pi_\mu = P_\mu - 
	\frac{1}{4 l^2}  K_\mu,
\label{TransiSbis}
\ee
as given by Eq.~(\ref{TransiS}).
The invariance of a source lagrangian under the transformations generated by $\Pi_\mu$ yields the conservation law \cite{dSgeod}
\be
\nabla_\mu \Pi^{\mu \nu} = 0
\label{dSconservation}
\ee
where
\be
\Pi^{\mu \nu} =
T^{\mu \nu} - \frac{1}{4l^2} \, {K}^{\mu \nu}
\label{TmK}
\ee
is the de Sitter-modified conserved current, with $T^{\mu \nu}$ the symmetric energy-momentum current, and
\be
K^{\mu \nu} =
\left(2 \eta_{\alpha \rho} \, x^\rho x^\nu -
\sigma^2 \delta_\alpha^\nu \right) T^{\mu \alpha}
\label{KdelT}
\ee
the proper conformal current~\cite{coleman}.

On the other hand, in the parameterisation appropriate to deal with small values of $l$, any two points of a locally-de Sitter spacetime separated by an infinitesimal distance will be connected to each other by a transformation generated by
\be
\bar \Pi_\mu = 4 l^2 P_\mu - K_\mu.
\label{TransiSbis2}
\ee
The invariance of a source lagrangian under such transformation yields the conservation law
\be
\bar \nabla_\mu \bar \Pi^{\mu \nu} = 0
\label{dSconservation2}
\ee
where
\be
\bar \Pi^{\mu \nu} =
4 l^2 T^{\mu \nu} - {K}^{\mu \nu}
\label{TmK2}
\ee
is the de Sitter-modified conserved current.

It is important to remark that the inclusion of the proper conformal transformations in the local spacetime transitivity---and consequently in the notions of space distance and time interval---does not change the number of the degrees of freedom. In fact, both Poincar\'e and de Sitter have ten degrees of freedom. However, there is a kind of {\em internal freedom} which is not present in locally Minkowski spacetimes. This freedom refers to the fact that now energy-momentum current can transform into proper conformal current, and vice versa, while keeping the total, or de Sitter notion of energy-momentum current covariantly conserved. As we are going to see, this possibility opens up important roads for the study of the cosmological evolution, and in particular for the study of cyclic models.

\section{Final remarks}

The de Sitter spacetime is usually interpreted as the simplest {\em dynamical} solution of the sourceless Einstein equation in the presence of a cosmological constant, standing on an equal footing with all other gravitational solutions---like for example Schwarzschild and Kerr. However, as a non-gravitational spacetime, the de Sitter solution should instead be interpreted as a fundamental background for the construction of physical theories, standing on an equal footing with the Minkowski solution. When general relativity is constructed on a de Sitter spacetime, since the cosmological term $\Lambda$ is now encoded in the local kinematics, it does not appear explicitly in Einstein equation. As an immediate consequence, in contrast to what happens in ordinary general relativity, the second Bianchi identity does not require $\Lambda$ to be constant \cite{new,hendrik}. It should be noted that the curvature tensor in this theory represents both the general relativity {\em dynamic curvature}, whose source is the energy-momentum current, and the {\em kinematic curvature} related to the underlying de Sitter special relativity.

In the de Sitter-modified general relativity, any solution to Einstein field equations will be a de Sitter-Cartan spacetime. In the contraction limit $l \to \infty$, corresponding to a vanishing cosmological term $\Lambda$, the underlying de Sitter spacetime contracts to Minkowski, and the de Sitter-Cartan geometry reduces to the usual Riemannian geometry of general relativity, which is consistent with the ordinary Poincar\'e-ruled special relativity. To study the contraction limit $l \to 0$, which corresponds to an infinite cosmological term $\Lambda$, one has to use the de Sitter ``translation'' generators as given by Eq.~(\ref{TransiSbis2}). In this limit, the translational degrees of freedom are switched off: only the proper conformal degrees of freedom remain active. Accordingly, the energy-momentum tensor is suppressed from the conserved current (\ref{TmK2}), which means that the proper conformal current turns out to be conserved. One can then say that the gravitational degrees of freedom are switched off in this limit, and consequently the Weyl curvature tensor must vanish. All matter content of the universe, therefore, must be in the form of proper conformal current---that is to say, it must be conformal invariant. This includes electromagnetic and gravitational waves,\footnote{Gravitational waves could in principle exist in the cone spacetime. There is a problem, however: the field equation for a symmetric second-rank tensor (perturbation of the metric) is not conformal invariant \cite{DesHen}. A possible solution to this puzzle is to reinterpret a spin 2 as a 1-form assuming values in the translation group (perturbation of the tetrad), in which case its field equation turns out to be conformal invariant \cite{s2CI}.} as well as any other conformal invariant fields. Furthermore, the general relativistic dynamic curvature vanishes, leaving only the kinematic de Sitter spacetime, which as we have already seen contracts to the four-dimensional, flat, singular, conic spacetime $\bar M$. Due to the fact that it is transitive under proper conformal transformations, its metric tensor $\bar{\eta}_{\mu \nu}$ does not have the conventional meaning of a physical metric: it is not dimensionless, and the interval it defines,
\be
d\bar{s}^2 = \sigma^{-4} \, \eta_{\mu \nu}\, dx^\mu dx^\nu,
\ee
has to do with the proper conformal notions of time interval and space distance. In such spacetime, therefore, the usual notions of space distance and time interval cannot be defined. In particular, our conventional (translational) notion of time does not exist on $\bar M$ \cite{tempo1,tempo2}. This means that local clocks cannot be defined, a result that is somehow in agreement with the general idea that (our conventional notion of) time should not exist at the Planck scale \cite{rovelli}. One should note, however, that the proper conformal notion of time does exist. Considering in addition that the metric on the conic spacetime $\bar{M}$ is conformal invariant, its singularity is a conformal gauge singularity, and it satisfies the Weyl curvature hypothesis, one can then consistently interpret $\bar{M}$ as the bridging spacetime connecting two aeons in Penrose's conformal cyclic cosmology \cite{PenBook}.

It should be remarked that $l \to 0$ is just a formal limit in the sense that quantum effects preclude it to be fully performed. It is actually a contraction limit, on an equal footing with the classical contraction limit in which the speed of light goes to infinity $c \to \infty$, as well as with other possible contraction limits \cite{contra1,contra2,contra3}. The cone spacetime $\bar{M}$, which emerges as the output of this limit, should then be thought of as a kind of {\em frozen geometric structure} behind the spacetime quantum fluctuations taking place at the Planck scale. The conic spacetime $\bar M$ can then be interpreted as a universe in which all energy content is in form of proper conformal current only. The quantum fluctuations from the cone spacetime with $l = 0$ to a de Sitter spacetime with $l = l_P$ give rise to a non-singular de Sitter universe with a huge cosmological term $\Lambda_P \simeq 10^{66}~\mbox{cm}^{-2}$, which could drive inflation. Once this transition occurs, the translational degrees of freedom are turned on, and our usual notions of time and space emerge, as can be seen from the generators (\ref{TransiSbis2}). At this point, however, owing to the tiny value of $l$, space and time will still be preponderantly determined by proper conformal transformations. Concomitantly, the proper conformal current begins transforming into energy-momentum current, giving rise to the baryonic matter present in the universe.

To finish, using cosmographic arguments, we present a speculative discussion on how a de Sitter-ruled special relativity can give rise to a cyclic view of the universe. As discussed above, at the Planck time $t_P$, the cosmological term assumes the huge value $\Lambda_P$, which gives rise to a rapidly expanding primordial universe. As the proper conformal current transforms into energy-momentum current, the cosmological term experiences a decaying process, leading to an era of decelerated expansion. This process continues until the cosmological term assumes a tiny value, in such a way to allow the formation of the cosmological structures we see today. In this period, most of the proper conformal current has already been transformed into energy-momentum current, and the de Sitter-Cartan structure of spacetime differs slightly from the usual Riemannian geometry. Accordingly, space distance and time interval in this period are preponderantly determined by ordinary translations.

Now, cosmological observations indicate that the universe has entered (a few billion years ago) an accelerated expansion era \cite{obs1,obs2,obs3}. From the point of view of a de Sitter-ruled special relativity, this means that the energy-momentum current began transforming back into proper conformal current, which produces an increase in the value of the cosmological term. Such process means that the universe has already begun moving back towards its initial state, characterised by a huge cosmological term---the final state of this era, as well as the initial state of the next era, represented by the cone spacetime $\bar M$. This picture, together with recent cosmological observations, point to a cyclic view of the universe, but with a completely new concept of a cyclic universe: as paradoxical as it may sound, this geometry gives rise to an ever expanding cyclic universe.

\section*{Acknowledgments}

The authors would like to thank an anonymous referee for valuable comments and suggestions. They would like to thank also FAPESP, CAPES and CNPq for partial financial support. A.A. thanks Universidad Centroccidental Lisandro Alvarado, Barquisimeto, Ve\-ne\-zue\-la, for financial support. 

\begin{appendix}

\section*{Appendix: On the notion of transitivity}

Spacetimes with constant sectional curvature are maximally symmetric in the sense that they can lodge the highest possible number of Killing vectors \cite{weinberg}. Their curvature tensor are completely specified by the scalar curvature $R$, which is constant throughout spacetime. Minkowski $M$, with vanishing curvature, is the simplest one. Its kinematic group is the Poinca\-r\'e group ${\mathcal P} = {\mathcal L} \oslash {\mathcal T}$, the semi-direct product of Lorentz (${\mathcal L}$) and the translation (${\mathcal T}$) groups. It is a homogeneous space under the Lorentz group:
\[
M = {\mathcal P}/{\mathcal L}.
\]
The Lorentz subgroup provides an isotropy around a given point of $M$, and the translation symmetry enforces this isotropy around any other point. This is the meaning of homogeneity: all points of Minkowski spacetime are equivalent under spacetime translations. That is to say, Minkowski is {transitive} under spacetime translations.

Another example of maximally symmetric spacetime is the de Sitter space $dS$. It has non-vanishing sectional curvature, and $SO(4,1)$ as kinematic group. Furthermore, it is also homogeneous under the Lorentz group:
\[
dS = SO(4,1) / {\mathcal L}.
\]
Like Minkowski, the Lorentz subgroup provides an isotropy around a given point of $dS$. The notion of homogeneity, however, is completely different: as one can see from the generators (\ref{TransiS}) or (\ref{TransGenL}), all points of the de Sitter spacetime are equivalent under a combination of translation and proper conformal transformation---the so-called de Sitter ``translations''. That is to say, de Sitter is transitive under a combination of translations and proper conformal transformations: in order to move from one point to any other point of a de Sitter spacetime, one has to perform a de Sitter ``translation''. This is the meaning of transitivity.
 
\end{appendix}


\end{document}